\documentclass[prb,twocolumn,superscriptaddress,prb]{revtex4}

\usepackage{amsmath,psfig,amssymb,graphicx,bm}

\renewcommand{\Re}{\mathop{\mathrm{Re}}}
\begin{document}
\title{Decoherence of a qubit by non-Gaussian noise 
at an arbitrary working point}
\author{J Bergli}
\email{jbergli@fys.uio.no}
\affiliation{Department of Physics and Center for Advanced Materials
  and Nanotechnology, University of Oslo, PO Box 1048
  Blindern, 0316 Oslo, Norway}
\affiliation{Physics Department, Princeton University, Princeton, NJ
  08544, USA} 
\author{Y. M. Galperin}
\affiliation{Department of Physics and Center for Advanced Materials
  and Nanotechnology, University of Oslo, PO Box 1048
  Blindern, 0316 Oslo, Norway}
\affiliation{Argonne National Laboratory, 9700 S. Cass Av.,
Argonne, IL 60439, USA}
\affiliation{A. F. Ioffe  Physico-Technical Institute of Russian Academy of
Sciences, 194021 St. Petersburg, Russia}
\author{B. L. Altshuler}
\affiliation{Physics Department, 
Princeton University, Princeton, NJ 08544, USA}

\begin{abstract}

The decoherence of a qubit due to a classical non-Gaussian noise with
correlation time longer than the decoherence time is discussed for
arbitrary working points of the qubit. A method is developed  that
allows an exact formula  for the phase memory functional in the
presence  of independent random telegraph noise sources to be derived.

\end{abstract}

\maketitle

\section{Introduction}
Reducing the decoherence induced by interaction with the environment is one of
the major challenges in the practical implementation of quantum
computing. In particular, for solid state qubits this seems to be the
most important issue. In this paper we discuss the effect of
noise with a correlation time that is long compared to the decoherence
time of the qubit.  The work is directly
motivated by experiments~\cite{Nakamura, Devoret} 
on Josephson charge qubits (JCQ), but due to their general nature
they can also be
relevant to many other systems. The JCQ is built around a
small superconducting grain connected to a superconducting reservoir
by a Josephson junction. By means of a capacitively coupled gate
voltage one can control the number of Cooper pairs on the small
grain. Because of the Coulomb blockade there will be a preferred
number of Cooper pairs except at special degeneracy points where the
energies of states with $n$ and $n+1$ coincide. The degeneracy is
lifted by the Josephson coupling, giving the usual level anticrossing
picture sketched in Fig.~\ref{f1}.
If the gate voltage is never too far from the degeneracy point we
can ignore transitions to other levels and the device can be
regarded as a two level system or qubit.

Consider noise in this system originating from fluctuations in the
electric potential of the grain. This could
be either due to fluctuations in the gate voltage source, or to fluctuating
charges in the environment. Both would correspond to fluctuations of
the working point position along the horizontal axis of the figure,
with a corresponding change 
in the energy.  The experiments\cite{Nakamura} where conducted
with the gate voltage away from the degeneracy point, where to a
good approximation the energy is a linear function of the
potential. This case is covered by the theory\cite{paladino,galperin} for the
case of free induction and narrow 
distribution of coupling constants and for both free induction and
echo at arbitrary distribution of coupling 
constants, respectively.

 Using a modified circuit Devoret \textit{et. al.}~\cite{Devoret}
were able to work at the degeneracy point where to first order
there is no change in energy when the potential is changed. The
idea is that this would make the device less sensitive to
electrostatic noise. This is then called
the \textit{optimal working point}. At this point the expectation value of
the charge is the same for both states. The purpose of this paper
is to extend the theory~\cite{paladino,galperin} to this situation and to
see how dephasing is changed as the optimal point is approached.

The paper is organized as follows. In section \ref{mod} the model for the 
qubit interacting with the noise is presented. In section \ref{st} we
discuss a short time  
expansion that illustrates in a simple  way the interplay 
of the effect of several uncorrelated fluctuators at the optimal point. 
The main purpose of this section is providing a qualitative
understanding, as the results are contained 
within the full solution presented in section \ref{fs}.

\section{Model}\label{mod}

\begin{figure}[h]
\centerline{ \psfig{file=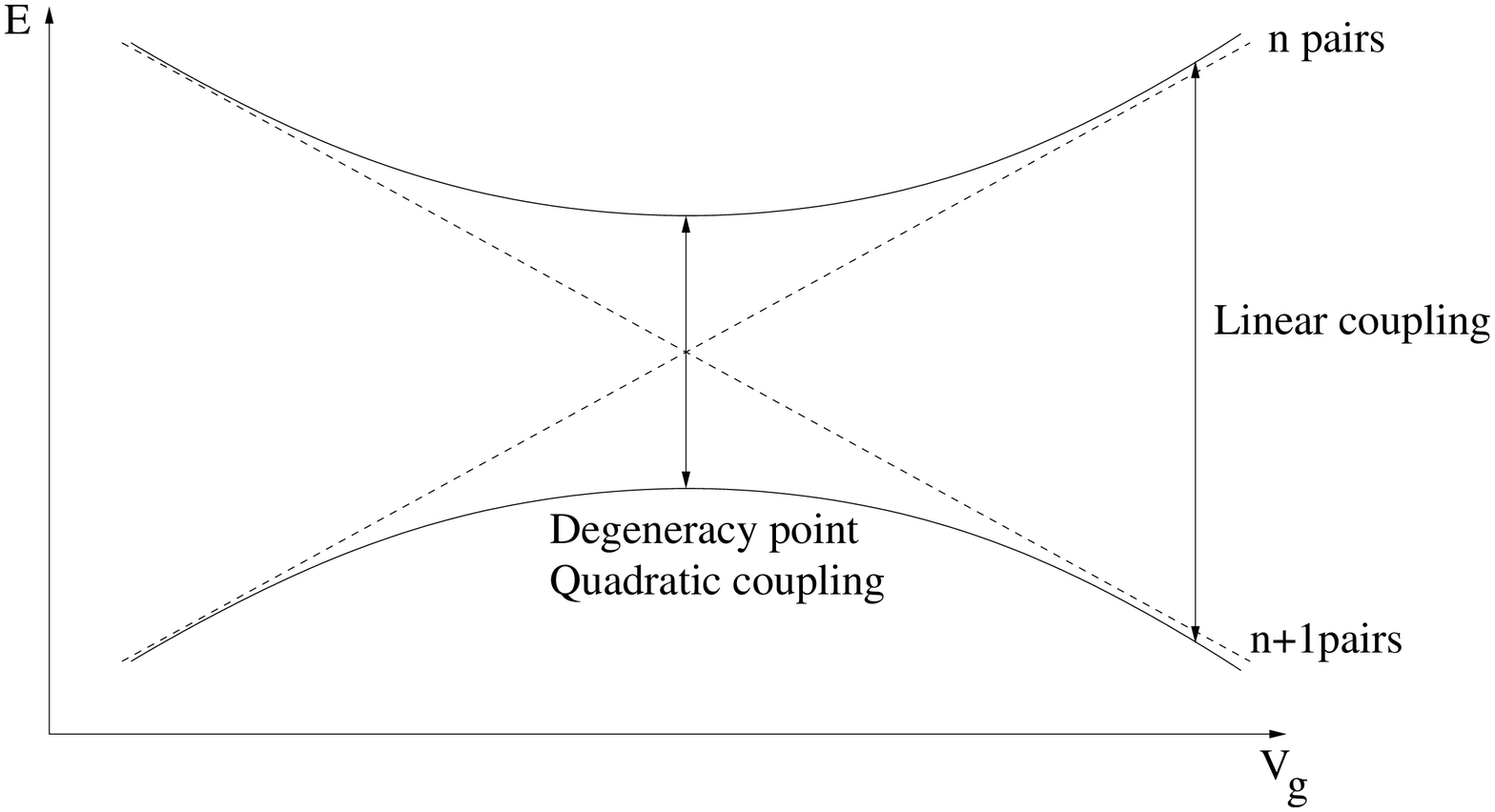,width=80mm,height=50mm}}
\caption{Energy levels of Josephson qubit as function of the gate
potential.} \label{f1}
\end{figure}

Consider the Hamiltonian of the Josephson qubit, c. f. with
Ref.~\onlinecite{makhlin-review},
\begin{equation}
H = \frac{1}{2}\Delta\, \sigma_z-\frac{1}{2}E_J\, \sigma_x\, .
\label{Ham}
\end{equation}
Here $\Delta=E_c C_g\left(V_g - V_g^{\text{opt}} \right)/e$,
 where $E_c$ is the charging energy, $\left(V_g -
 V_g^{\text{opt}}\right)$ is the deviation in gate 
voltage from the optimal working point, $C_g$ is the gate
capacitance and $e$ the electron charge. The energy levels as function
of the gate potential are shown in Fig.~\ref{f1}.
In principle, both $\Delta$ and $E_J$ can fluctuate in time, but for
clarity we will  only consider the noise in $\Delta$. Noise in $E_J$
was considered in Ref.~\onlinecite{vanHarlingen} and can be treated
in a similar way.  

We will describe fluctuations in $\Delta$ by a random time-dependent quantity
$v(t)$. The Hamiltonian of the qubit
is then
\begin{equation}
H = \frac{1}{2}[\Delta+v(t)]\sigma_z-\frac{1}{2}E_J \sigma_x\, .
\label{Ham0}
\end{equation}
One can express $v(t)$ through fluctuations of the effective
gate voltage, $\delta V_g (t)$ as $v(t)=E_c C_g\,
\delta V_g(t)/e$. One of the sources of such fluctuations is hopping
background charges.\cite{paladino,galperin,makhlin} 
We briefly recall the model of Refs.~\onlinecite{paladino,galperin}.
It is assumed that there are dynamic defects characterized by two
metastable states. They can be, for example, traps near the
electrodes able to capture electrons from the electrodes and then
re-emit them, or pairs of traps with one electron bouncing between them.
Each dynamic defect is represented as a classical fluctuator producing
random telegraph noise. That is, the state of each fluctuator is
represented by a random function $\chi(t)$ which switches between the two
values, $\pm 1/2$. The probability to switch $n$ times between theses
states during the time $t$ is assumed to be given by the Poisson distribution 
\begin{equation}
  \label{eq:pd1}
  P_n(t)= \frac{(\gamma t)^n}{n!}e^{-\gamma t}\, .
\end{equation}
 Here $\gamma$ is the characteristic switching frequency of the given
 fluctuator. This means that the switching events occur independently
of each other. 
The fluctuation $v(t)$ is a sum  of the contributions of different
fluctuators,  
\begin{equation}
  \label{eq:pert1}
 v(t) =\sum_i v_i\chi_i(t)\,. 
\end{equation}
Different fluctuators are assumed to be statistically independent, 
\begin{equation}\label{corr}
 \langle\chi_i(t_1)\chi_k(t_2)\rangle =
 \frac{1}{4}e^{-2\gamma_i|t_1-t_2|} \delta_{ik} \, .
\end{equation}
This is actually the simplest model since it is assumed that the
probabilities to jump between the states in two directions are the
same. As a result, equilibrium populations of the states are
equal. This assumption can be relaxed~\cite{galperin} without change
of the physical conclusions. Indeed, since only fluctuators with 
energy splittings $\lesssim kT$ contribute to the decoherence, the
equilibrium populations of the efficient states do not essentially differ. 

Several papers~\cite{galperin,paladino,makhlin} have addressed the
dephasing in the presence of a large number of fluctuators with a
broad range of switching rates, $\gamma$. This situation is
relevant for systems showing $1/f$ noise. We believe, however,
that the experiments with ``good" qubits, e. g.,
Ref.~\onlinecite{Nakamura},  are best understood in terms of a
small number of fluctuators.
Also, if one is to achieve quantum computation this is
necessary as a large number of fluctuators most probably will lead
to large decoherence and ruin the device. We will therefore here focus
on the effect of a small number of fluctuators.

The noise affects the qubit in two ways: (i) it causes shifts in the 
energy levels of the two states and thereby introduces a random contribution 
to the relative phase of the two states (dephasing) and (ii) it causes 
transitions between the two states leading to energy relaxation. Let
us for the moment concentrate on the  
first effect, dephasing. 
The important quantity that we study is the phase-memory functional
\begin{equation}\label{pmf}
  \Psi = \left\langle \exp \left[ \frac{i}{\hbar}\int_0^tdt'\beta (t')\delta E(\{\chi_i(t')\})\right] \right\rangle
\end{equation}
where $\delta E(\{\chi_i(t')\})$ is the shift in the energy
splitting of the two levels caused by the fluctuators, while
$\beta (t)$ is some function which depends on the 
qubit manipulation procedure, as will be explained below. 
The phase-memory functional describes
the relative  phase picked up during time evolution by one state of the qubit
relative to the other.

Diagonalizing Eq.~(\ref{Ham0}) we get the eigenenergies
\begin{equation}\label{energies}
  E_\pm = \pm\frac{1}{2}\sqrt{(\Delta +v )^2+E_J^2} \, .
\end{equation}
If the working point is far from the optimal one, $\Delta \gg E_J$,
we can neglect the Josephson 
energy, and coupling of the fluctuators to the
qubit is linear,
\begin{equation}
E_\pm \approx \pm \frac{1}{2}\sqrt{\Delta^2+E_J^2}\left( 1 +\frac{v
  \Delta}{\Delta^2+E_J^2}\right) \, .  
\end{equation}
An exact formula for the phase memory functional for a linear coupling in the case of a single fluctuator coupled to the qubit was derived in
Refs.~\onlinecite{galperin,paladino,lerner}. In this
limit, because of the linearity $E$ in $v$, the phase memory
functional for any number of fluctuators is found by simply
multiplying the phase memory functionals of different fluctuators. 
In the regime of exponential decay this procedure corresponds to
simply adding the decay rates and the resulting decay rate is
represented by the average over the distributions of fluctuator
parameters.\cite{galperin} The optimal point, $\Delta=0$, was studied  
in various approximations and numerically in
Refs.~\onlinecite{makhlin,makhlin2}. 

The aim of the present work is to derive an exact formula for 
the phase memory functional of a single fluctuator similar to that in
Refs.~\onlinecite{galperin,paladino} that is applicable at an
\textit{arbitrary} working point.   
Averaging over a large number of fluctuators is in the general case not 
as simple as in the linear regime since the phase memory functional is not 
the product of individual factors for each fluctuator. However, we
will extend the analysis to a small number of fluctuators, which we believe is 
relevant to qubit experiments.\cite{Nakamura}

To make the main idea clear let us first study the expansion at short
times, $\gamma t \ll 1$. Firstly, this expansion provides some
insight; secondly, the short-time situation can be
most important for realistic qubits since at long times the phase
memory functional has already decayed to a very low value. 

\section{Phase memory functional  at small times, $\gamma t \ll 1$}\label{st}

We expand the energy in the limit 
$$v \ll E_0  \equiv \sqrt{\Delta^2+E_J^2}$$ to 
obtain
\begin{equation}
  E_\pm = \pm \frac{1}{2}\left(E_0
  +\frac{\Delta \, v}{E_0} +\frac{v^2}{2E^*}\right)
 \end{equation}
where $E^* \equiv E_0^{3}/E_J^2$.
When averaging the phases we subtract the initial values so that
we only get what comes from fluctuator jumps. Thus $\Psi \equiv
\left< e^{i\phi(t)}\right>$ where
\begin{eqnarray} \label{phi-t}
  \phi(t) &=& \frac{1}{\hbar}\int_0^t \beta(t')
  dt'\left[\sum_i\frac{\Delta}{2E_0}\, v_i[\chi_i (t')-\chi_i^0]
  \right.  \nonumber \\  && \left. +
       \sum_{ij}\frac{v_iv_j}{4E^*}\left[\chi_i(t')\chi_j(t')-\chi_i^0\chi_j^0\right]\right] 
\end{eqnarray}
is the random phase shift. Here $\chi_i^0 \equiv \chi_i(0)$ while the
average is calculated over random telegraph processes in the fluctuators.

If there is only linear coupling, each fluctuator appears in only
one term in the exponent. Since we assume the fluctuators to
be statistically independent the average can be split into a product
of averages 
over each individual fluctuator. With the quadratic term included
this is no longer possible. Here we will study the cases of
one and two fluctuators. This gives some insight to the general
structure, and it is also the most relevant for realizing a
working qubit since a large number of fluctuators would destroy
the working of the qubit. 

\paragraph[*]{Free induction signal:}
Let us  assume that $\beta (t)=\theta (t)$ where
$\theta (t)$ is the Heaviside unit step function. This assumption
corresponds to the free induction signal. With only one
fluctuator the quadratic term is identically zero, and we have
\begin{equation}
  \Psi_1=\left\langle e^{i\nu \int_0^t
  dt'[\chi(t')-\chi^0]}\right\rangle\, , \quad \nu \equiv
  \frac{\Delta}{2E_0 \hbar  }\, v\, .
\end{equation}
Let us evaluate the memory functional in the limit of $\gamma t\ll 1$,
first for one, and then for two fluctuators. Then the
possibility of more than one jump is negligible, and we write for
the probabilities of zero and one jump $P_0=1-\gamma t$ and
$P_1=\gamma t$. The memory functional can be calculated by averaging
over the time, $t_1$,  between the jumps, which at $\gamma t \ll 1$
are equally probable:  
\[ 
  \Psi_1 = P_0 + \frac{P_1}{t}\int_0^t \! \! dt_1 \cos \nu (t-t_1)=
  1-\frac{\gamma }{\nu} (\nu  t-\sin\nu t)\, .
\] 
If
$
 \nu t=(\Delta/2E_J\hbar)v t\ll 1
$
 we can expand the sine and
get $\Psi_1=1-\gamma\nu^2t^3/6$. At the optimal point $\nu=0$ and the
fluctuator is not visible to the qubit. At other working points
the dephasing rate is proportional to $\Delta^2$.

Now take the case of two fluctuators. We have to calculate
$\Psi_2 \equiv \left< e^{i\phi_2(t)}\right>$ where
\begin{eqnarray}
  \phi_2 (t)&=&\int_0^t dt'
  \left[\nu_1(\chi_1-\chi_1^0)+\nu_2(\chi_2-\chi_2^0)
  \right. \nonumber \\&& \left. \qquad
    + 2\lambda_{12}(\chi_1\chi_2-\chi_1^0\chi_2^0)\right]
    \label{phi2-t}
\end{eqnarray}
and $\lambda_{12}=v_1v_2/E^* \hbar$. Again, for $\gamma_i  t\ll 1$ we
get
\begin{equation}
\Psi_2=1-\sum_{\pm,i=1,2 } \left[\gamma_i t -
\frac{\gamma_i}{2}\left(\frac{\sin(\nu_i
\pm\lambda_{12})t}{\nu_i\pm \lambda_{12}}
             \right)\right]\, .
\end{equation}
Assuming $\nu_1,\nu_2,\lambda_{12}\ll t^{-1}$ and expanding
sines we get:
\begin{equation}
\begin{split}
  \Psi_2&=1-\sum_i \frac{\gamma_i}{6}(\nu_i^2+\lambda_{12}^2)t^3 \\
   &\approx
   \left(1-\frac{\gamma_1+\gamma_2}{6}\lambda_{12}^2t^3\right)
   \prod_i \left(1-\frac{\gamma_i\nu_i^2t^3}{6}\right)\, .
\end{split}
\end{equation}
We see that in this limit one can split the expression for $\Psi$
in factors corresponding to the individual fluctuators
just as in
the case of linear coupling, but there appears an additional factor 
due to the nonlinear coupling. 
 It is easy to see that a similar pattern will also appear for larger
 number of fluctuators. At longer times this product structure 
is lost. 

The interplay between the linear and quadratic coupling is now
quite clear. If $\nu\gg \lambda$ the linear coupling is
dominant. Approaching the optimal point will reduce the dephasing
until $\nu$ becomes smaller than $\lambda$ where the term
proportional to $\lambda^2$ becomes most important. This
contribution results from the interplay between the two
fluctuators and cannot be eliminated; thus it represents the
minimal dephasing possible at the optimal point. The physical
reason for this is quite easy to understand. Switching of
fluctuator 1 shifts the average point that fluctuator 2 is working
around. Both positions of fluctuator 1 can not represent the
optimal point with respect to fluctuator 2, and some dephasing is bound to
occur. This is the most important physical insight that
distinguishes the quadratic coupling from the linear. In the case
of quadratic coupling, even if the different fluctuators in
themselves are independent, their effect on the qubit will be
influenced by the positions of all the others. With linear
coupling one finds that slow fluctuators, with $\gamma t \ll1$ do
not contribute to the dephasing. This is no longer true for
quadratic coupling, as they play a role in determining the effect
of the fast fluctuators even if they do not have time to switch
during the experiment. Thus very slow fluctuators may be of great
importance.

\paragraph[*]{Two-pulse echo:}
Perhaps more directly related to experiments are the echo signals.
These are found using Eq.~(\ref{phi-t}) where for two-pulse echo
\begin{equation}
\beta(t)=\left\{ \begin{array}{lcl} +1 &\text{for}& t<\tau \, ,\\
-1 & \text{for}& \tau<t<2\tau\, . \end{array} \right.
\end{equation}
Here $\tau$ is the delay between the initial
pulse and the echo pulse, and the echo signal is centered around
$2\tau$. For short times ($\gamma\tau<1$) this gives for one
fluctuator
\begin{equation}
    \Psi_1= 1-\frac{2\gamma}{\nu}(\nu \tau -\sin\nu
    \tau)\, ,
\end{equation}
and for two fluctuators
\begin{equation}
\Psi_2=1-\sum_{\pm,i=1,2 } \left[2 \gamma_i \tau -
\gamma_i\left(\frac{\sin(\nu_i
\pm\lambda_{12})\tau}{\nu_i\pm \lambda_{12}}
             \right)\right]\, .
\end{equation}

\section{Exact solution for small number of fluctuators}\label{fs}

We now turn to the calculation of the phase memory functional for
arbitrary times.
The method described here is in principle applicable to an arbitrary
number of fluctuators, but the resulting formulas quickly get
impracticably  large when the number of fluctuators increase.  The main
idea is to consider the phase $\phi$ as a random variable with some
probability distribution $p(\phi,t)$ that will depend on time. Once
this is known the phase memory-function is given by $\Psi = \int dx\,
e^{i\phi} p(\phi,t)$. By mapping to a correlated random walk problem we
derive a Master equation for the probabilities $p(\phi,t)$.  The
details of how to calculate $p(\phi,t)$ are given in
appendix~\ref{a1}.

\subsection{Distribution function for one fluctuator far from the optimal point}

\begin{figure}[h]
\centerline{ 
\includegraphics[width=\columnwidth]{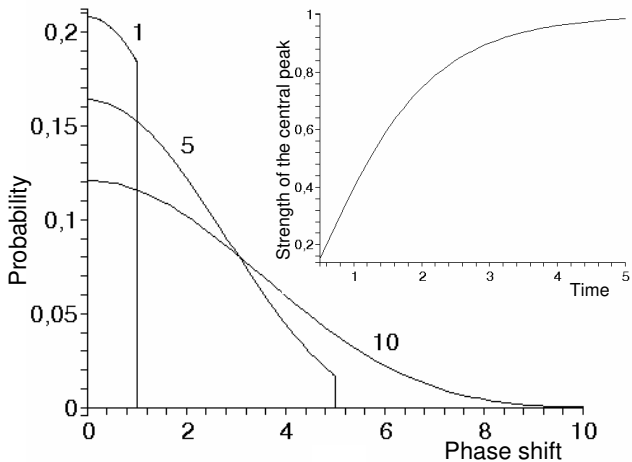}
}
\caption{Distribution function at time 1, 5, 10 for $\gamma=1$ and
  $\nu =1$. Inset: time dependence of the integrated strength of the
  central peak.}
\label{fordeling}
\end{figure}

\paragraph*{Free-induction signal:}
Let us first discuss the results for the distribution function in the
simplest case, that of one fluctuator far from the optimal point. The phase 
memory functional for this problem was derived in
Refs.~\onlinecite{paladino,galperin},
and in the end we will rederive the same expression. However, it is the 
simplest example for illustrating the general method, and it gives new 
insight to the relation between the Gaussian approximation and the 
fluctuator model.  To
better understand the meaning of the results it is useful to recall
the standard picture of dephasing by a \textit{Gaussian} noise that is
well known from NMR-physics (see, e. g., Ref.~\onlinecite{NMRbok}). If
the time $t$ entering the phase memory functional
\[
 \Psi = \langle e^{i\phi (t)}\rangle, \quad \phi (t)= \int_0^t dt' \nu (t')
 \, , \quad \nu(t) \equiv \nu\, \chi (t)
\]
is much longer than the correlation time $\approx \gamma^{-1}$ of the 
fluctuating function $\nu (t)$,  the integral can be considered as the sum 
of a large number of uncorrelated contributions. Consequently, by the 
central limit theorem, the phase will be distributed according to 
a Gaussian 
\[
 p(\phi) = \frac{1}{\sqrt{2\pi\langle\phi^2\rangle}}e^{-\frac{\phi^2}
   {2\langle\phi^2\rangle}}
\]
and the phase memory function is
\begin{equation}\label{pmg}
\Psi = e^{-\langle\phi^2\rangle/2} \, .
\end{equation}
From Eq.~(\ref{corr}) for the correlation function  we get
\[
 \langle\phi^2\rangle = \frac{\nu^2}{4\gamma}t+\frac{\nu^2}{2\gamma^2}
\left(e^{-2\gamma t}-1\right)
\approx 
\frac{\nu^2}{4\gamma}t \ \text{at}\ \gamma t \gg 1\, .
\]
Thus, in the Gaussian approximation,  the phase memory functional
 decays  exponentially at $\gamma t\gg1$ with the rate 
\begin{equation}\label{g2g}
\Gamma_\phi^{(G)}=\nu^2/8\gamma \, .
\end{equation}
Using the method explained in Appendix \ref{a1} we can find an exact
 solution for the distribution function of $\phi$
\begin{widetext}
  \begin{equation}
    \label{eq:df001}
  p(\phi,t) = e^{-\gamma t} \left[\delta\left(\phi\mp \nu t/2\right)
    +\frac{\gamma}{\nu} \, \frac{( t\pm 2\phi/\nu )I_1\left(\gamma\sqrt{
    t^2-(2\phi/\nu )^2}\right)}
    {\sqrt{t^2-(2\phi /\nu)^2}}\right]\left[\theta\left(\frac{2\phi} {\nu}+
    t\right)-\theta\left(\frac{2\phi}{\nu}- t\right)\right]
  \end{equation}
 \end{widetext}
where the different signs correspond to different initial states of the 
fluctuator and $I_1(z)$ is the modified Bessel function.
Without jumps of the fluctuator, the result would be only the moving
$\delta$-pulse of a constant amplitude, $\delta\left(\phi\mp
  vt/2\right)$. The value $vt/2$ is 
the maximal possible value of $\phi$ acquired for the time $t$, while
the jumps of  the fluctuators account for the smooth part. 
Averaging over the initial  
state of the fluctuator we get (not writing the $\theta$-functions)
\begin{eqnarray}\label{df}
p(\phi,t) &=& e^{-\gamma t} \left[\frac{\delta(\phi-
    \nu t/2)+\delta(\phi+ \nu t/2)}{2} \right. \nonumber \\
&&+\left. \frac{\gamma}{\nu} \cdot \frac{I_1(\gamma t\sqrt{1-(2/\nu t)^2
    \phi^2})} {\sqrt{1-(2/\nu t)^2\phi^2}}\right]\, .
\end{eqnarray}
This is plotted in 
Fig.~\ref{fordeling} for the times $t=1,5,10$, $\gamma=1$ and $\nu =1$.
 We observe that the central region is similar to a Gaussian, 
but at short times this is cut off by the $\delta$-functions represented 
by the vertical lines.
At $\gamma t\gg1$ the function is indeed close to a Gaussian, as
can be seen from the asymptotic expansion of the Bessel function
\[
e^{-\gamma t} \frac{I_1(\gamma t\sqrt{1-(2/\nu t)^2
    \phi^2})} {2\sqrt{1-(2/\nu t)^2\phi^2}}
    \sim \frac{1}{\sqrt{2\pi\gamma t}} e^{-\frac{2\gamma\phi^2}{\nu ^2t}}
\]
Comparing to Eq.~(\ref{pmg}) we see that we recover the result for the 
dephasing rate in Eq.~(\ref{g2g}).

 However, we know from Ref.~\onlinecite{galperin}
that if $\nu >2\gamma$ we have pronounced non-Gaussian behavior. We can now 
understand this from the point of view of the distribution function.
The smooth central part of this indeed approaches a Gaussian for 
$\gamma t\gg1$ and this gives the dephasing rate (\ref{g2g}), but the 
$\delta$-functions at the ends only decay at the rate $\gamma$. As long as 
the $\Gamma_\phi^{(G)}$ of (\ref{g2g}) is smaller than $\gamma$ the
decay will be  
controlled by the central part and the Gaussian approximation is valid. 
If $\Gamma_\phi^{(G)} >\gamma$ the decay is limited by the
$\delta$-functions, and  
is set by the rate $\gamma$. 

\begin{figure}[h]
\centerline{
\includegraphics[width=7cm]{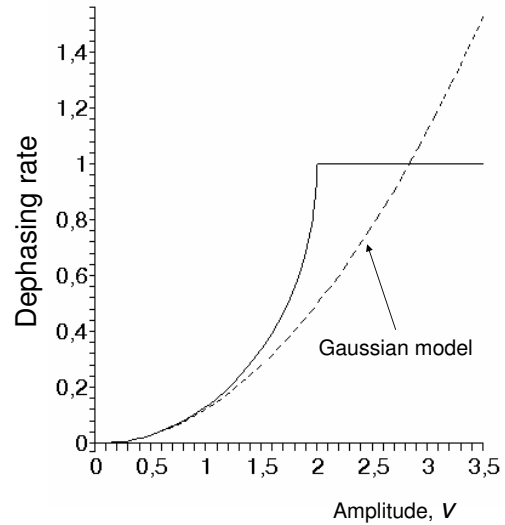}
}
\caption{Dephasing rate as function of $\nu$, $\gamma=1$. The solid line is 
the exact result for one fluctuator, Eq. (\ref{eq:dt21}), and the dashed 
line is the Gaussian approximation, Eq. (\ref{g2g}). } \label{G2}
\end{figure}

\begin{figure}[ht]
\centerline{
\includegraphics[width=7cm]{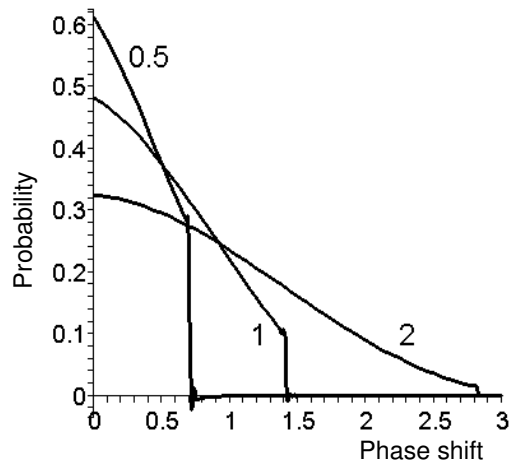}
}
\caption{Distribution of the phases for the echo signal for
  $\nu=\gamma$ and different products $\gamma \tau$ (shown near the
  curves). \label{fig:echo01}}
\end{figure}

Using the distribution function (\ref{df}) one can calculate the 
phase memory functional (see Appendix \ref{a1}) and one finds for 
$\gamma t\ll1$ an exponential decay with rate
\begin{equation}
  \label{eq:dt21}
  \Gamma_\phi=\gamma - \Re \left(\sqrt{\gamma^2-\nu^2/4} \right) \, .
\end{equation}
Figure~\ref{G2} shows $\Gamma_\phi$ as function 
of $\nu$ at $\gamma=1$ for the Gaussian and the fluctuator models.
\paragraph*{Two-pulse echo:} The same method as outlined in the
Appendix~\ref{a1} allows one to calculate the phase distribution
function, $p_e(\phi,\tau)$
for the two-pulse echo signal. The result can be expressed in the form
\begin{eqnarray}
  \label{eq:echo01}
 && p_e(\phi,\tau)=e^{-2\gamma \tau}\left[ \delta(\phi)
    \phantom{\int_0^\infty}\right. \nonumber \\
&&\left. +\frac{\gamma}{\pi} \int_0^\infty \! \! \! \! dk\frac{ \cos
  k\phi \sin w_k \tau \left(\gamma
  \sin w_k \tau
  +w_k  \cos w_k\tau\right) }{w_k^2}\right]\, . \nonumber
\end{eqnarray}
Here $w_k \equiv  \sqrt{(k\nu/2)^2-\gamma^2}$, while $\tau$ is the
delay time between the first and second pulse. The smooth part of the
distribution given by the second item in the above formula is plotted
in Fig.~\ref{fig:echo01}.
It is qualitatively similar to the phase distribution for the free
induction. 
\subsection{Dephasing rate as function of working point}
Let us then consider the decay of the phase memory functional
\textit{for different  
working points} . Again we recall the situation in NMR-physics where the loss
of the signal after the spins are set precessing by  a $\pi/2$-pulse is 
caused by two independent processes. The phase memory functional considered
above measures the random contributions
to the phase caused by fluctuating energy difference between the two states 
(which in an NMR experiment is caused by fluctuations in the magnetic field 
parallel to the external main field). In addition there are processes which 
flip the spin from one state to another, so called $T_1$-processes. These 
are caused by fluctuations in the magnetic field normal to the external field. 
If we denote the decay rate of the excited state into the ground state 
$\Gamma_1$ and add the two contributions we have the total decay of the
spin precession signal
\begin{equation}\label{g2}
 \Gamma_2 = \frac{1}{2}\Gamma_1 + \Gamma_\phi \, .
\end{equation}
The factor $1/2$ in front of $\Gamma_1$ can be understood from the fact that 
if the probability of the excited  state decays with rate $\Gamma_1$, the 
amplitude decays with the rate $\Gamma_1/2$ and this is what enters the 
off-diagonal elements of the density matrix. For a more detailed discussion 
see Ref.~\onlinecite{NMRbok}, or, for qubits, Ref.~\onlinecite{makhlin2}.

\begin{figure}[h]
\centerline{ \psfig{file=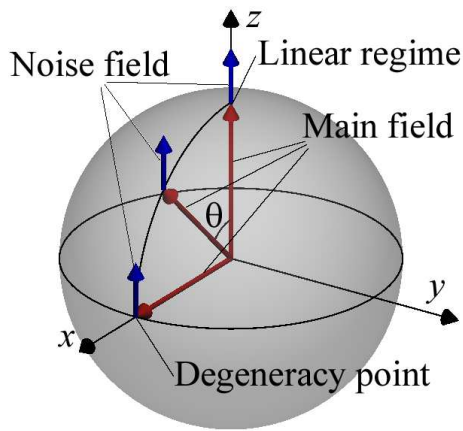,width=80mm,height=60mm}}
\caption{Different working points represented on the 
Bloch sphere} \label{bloch}
\end{figure}

Let us  now discuss how the relative strength of the two terms of Eq.~(\ref{g2}) 
changes as we change the working point of the qubit. 
Looking back at the Hamiltonian (\ref{Ham0}) we remember that it is 
equivalent to a spin 1/2-particle in a static magnetic field 
${\bf B} = E_J{\bf e}_x +\Delta{\bf e}_z$ while the noise is always
along the $z$-axis, $\bm{ \nu } = \nu {\bf e}_z$.
We denote the angle between ${\bf B}$ and the $z$-axis by  $\theta=
\arctan(E_J/\Delta)$.
In particular, $\theta=0$ corresponds to working far from the 
degeneracy point where $\delta \gg E_J$ while $\theta=\pi/2$ is the
degeneracy (optimal) point $\Delta=0$.
The time evolution of the qubit is then a precession on the 
Bloch sphere around the total field (Fig.~\ref{bloch}). 
Far from the degeneracy point the main field is directed along the $z$-axis
(pointing at the  
north pole of the sphere), whereas at degeneracy it is along the
$x$-axis (on the equator). All the time the noise vector $v$ is in the
$z$-direction. We see that the noise component parallel to the external 
field, which gives the $\Gamma_\phi$, is largest far from degeneracy, which 
agrees with our previous discussion. However, the noise normal to the 
field, giving $T_1$-processes, is maximal at degeneracy. 
In the Gaussian approximation this is given by (see
Ref.~\onlinecite{NMRbok}) 
 \[
 \Gamma_1 \equiv \frac{1}{T_1}=\sin^2\theta\int_0^\infty
   d\tau\,\langle \nu (t) \nu (t+\tau)   \rangle \, \cos \frac{E_0 \tau }{\hbar}\, .
\]
Using the correlator (\ref{corr}) one can rewrite this expression for
the case of $N$ identical fluctuators with parameters $\nu$ and
$\gamma$ as 
\begin{equation}\label{g1g}
\Gamma_1 =\frac{ N}{2}\,  \frac{\hbar^2\nu^2\gamma }{E_0^2+4\gamma^2}\,  \left(\frac{E_J}{E_0} \right)^2 \
    \, . 
\end{equation}
The Gaussian approximation for the $\Gamma_\phi$ is more difficult to obtain 
because the square root in the energy (\ref{energies}) makes the 
average in Eq.~(\ref{pmg}) not treatable analytically. However we can
expand in the  
lowest order in $v/E_0$, which is a good approximation except a close
vicinity of the degeneracy point, where the coefficient in front goes
to zero and higher  
order terms need to be calculated. This gives the same result as the 
$\Gamma_2^{\text{ad}}=1/T_2^{\text{ad}}$ of Ref.~\onlinecite{NMRbok}, 
\[
 \Gamma_\phi^{(G)} \approx
 \frac{1}{2}\cos^2\theta\int_0^\infty d\tau\langle \nu (t) \nu (t+\tau)
   \rangle \, ,
\]
which for $N$ identical uncorrelated fluctuators yields
\begin{equation}\label{g2a}
\Gamma_\phi^{(G)}  = \frac{N}{8}\, \frac{\nu^2}{\gamma} \,  
  \left(\frac{\Delta}{E_0}\right)^2 \, .
\end{equation}
Now we want to compare these Gaussian results with the exact expressions 
found using the method of Appendix \ref{a1}. 
\begin{figure}[h]
\centerline{\includegraphics[width=\columnwidth]{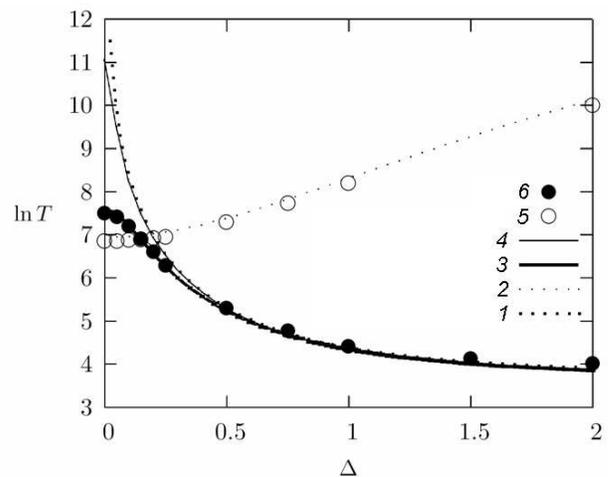}}
\caption{Relaxation times for the case of two fluctuators as function
  of working point for weak  coupling, $v/E_J=0.1$, and
  $\gamma/E_J=0.1$.  $\Delta$ is measured in units of $E_J$. 1 -- $1/\Gamma_\phi^{(G)}$, Eq.~(\ref{g2a}), 2 -- $T_1$, Eq.~(\ref{g1g}), 
  3 -- $T_2\equiv (1/2T_1 + 1/T_\phi)^{-1}$, 4 -- $T_\phi$, calculated
  according to exact expressions from  Appendix~\ref{a1}. 5, 6 -- results of
  numerical simulation for $T_1$ and $T_2$, respectively}
\label{f3} 
\end{figure}
In Figure~\ref{f3}  the relaxation times (inverses of the
relaxation rates) $T_1$ (curve 2) and $1/\Gamma_\phi^{(G)}$ (curve 1) 
are shown. They are
calculated according to Eqs.~(\ref{g1g}) and 
(\ref{g2a}), respectively.  The decay time, $T_\phi=\Gamma_\phi^{-1}$ (curve 4), 
of the phase memory functional calculated using the method of
Appendix~\ref{a1} and  
the resulting decay time, $T_2$ (curve 3), of the spin signal according to
Eq.~(\ref{g2}) are also shown. The points represent the rates obtained
by a numerical  simulation of the time evolution according to the Hamiltonian
(\ref{Ham0}), which performs averaging over many realizations of the random
process. All curves are calculated for the case of 2 fluctuators with 
coupling strength $\nu/E_J=0.1$ and switching rate
$\gamma/E_J=0.1$. Thus $v/\gamma = 1$ and 
this case belongs to the so-called weak
coupling regime.~\cite{galperin} 
We see that for all working points the decay is well 
described by the Gaussian approximation. This is because the rates
always are slower than the limiting rate $\gamma$ set by the
correlation time of the fluctuators, similar to what was described
above for the case far from the degeneracy point. For most working
points the rate $\Gamma_\phi\gg\Gamma_1$ and this dominates the 
$\Gamma_2$. Close to the degeneracy point we see that $\Gamma_1$ becomes
more important and at degeneracy it dominates completely giving
$\Gamma_2\approx\Gamma_1/2$. 
\begin{figure}[h]
\centerline{\includegraphics[width=\columnwidth]{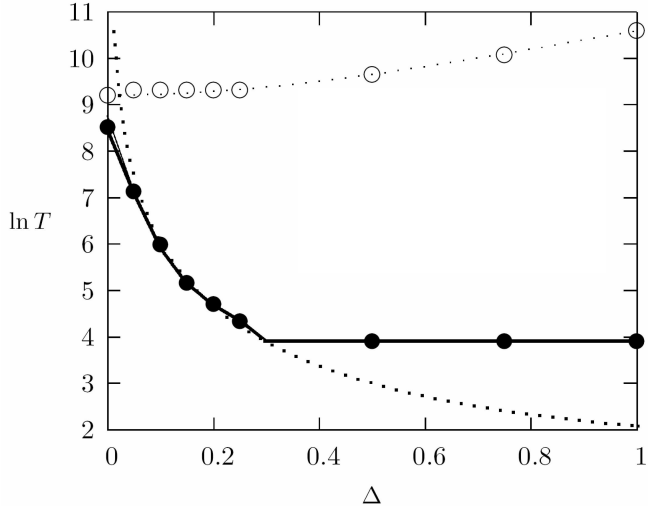}}
\caption{Relaxation times as function of working point for the case of
  strong coupling. Two fluctuators with $v/E_J=0.1$,
  $\gamma/E_J=0.01$. The legend is the same as in Fig.~\ref{f3}} \label{f4} 
\end{figure}
The same quantities for the case of strong coupling 
$v/E_J=0.1, \ \gamma/E_j=0.01$, $v/\gamma = 10$ are shown in Figure~\ref{f4}. 
Far from the degeneracy the
situation is similar to the one described earlier, with the rate,
$\gamma$,  determined
by the decay of the $\delta$-function peaks in the distribution
function. Closer to the degeneracy the rate is
slower and the Gaussian approximation gives good results.

 Note that in the case of strong coupling the difference between the
 approximate expression $\Gamma_\phi^{(G)}$ and the exact $\Gamma_\phi$
 becomes more noticeable.   
At large $\Delta$ this is because of the essentially non-Gaussian
 character of the noise, as 
discussed before. At small $\Delta$ the difference comes from the fact
that the $\Gamma_\phi^{(G)}$ only gives the Gaussian approximation to
$\Gamma_\phi$ in the lowest order in $\nu/E_J$. We expect that if we had
calculated the phase memory function in the Gaussian approximation 
according to (\ref{pmg}) and using the exact expectation value 
$\langle\phi^2\rangle$ (which is hard to find analytically) the result would 
agree completely with the $\Gamma_\phi$ calculated by the method of 
Appendix~\ref{a1} since the decay time is much longer than the 
correlation time $1/\gamma$ of the noise and the central limit theorem 
should work. Note also that at degeneracy the decay is still dominated 
by the phase relaxation processes, $\Gamma_\phi$, while the 
$T_1$-processes only give a small correction.

Figures~\ref{f3} and~\ref{f4}  should be compared to the experimental results of
Astafiev \textit{et al.}~\cite{astafiev2}  There is clear qualitative
agreement, but it is hard to try to make a quantitative fit,
especially for the $T_2$ where there is very little data. 

Let us
rather look back at the Hamiltonian (\ref{Ham0}) and
Figure~\ref{bloch}.  
So far we only considered noise in the $\sigma_z$ 
component of the Hamiltonian, as appropriate for noise sources coupled
to the charge of the qubit. As discussed in
Ref.~\onlinecite{vanHarlingen} there is also the possibility of noise
in the Josephson coupling (the $\sigma_x$ part). In Figure~\ref{bloch}
this would correspond to a noise vector parallel to 
the $x$-axis. This noise would be transversal far from degeneracy, giving 
large $T_2$ and small $T_1$ and it would be longitudinal at degeneracy
point, 
with $T_1$ large and $T_2$ small. The fact that the experimental results 
are similar to our figures~\ref{f3} and \ref{f4} rather that the opposite 
shows that in these experiments the noise in the $\sigma_z$ part is dominant.

\section{Discussion}
At the end we would like to discuss two simple observations based on the 
above results.
 
\subsection{Measuring the distribution function?}

The method that we have used to find the phase memory functional gave
as an intermediate result an expression for the distribution function
of the phase $p(\phi,t)$. The phase memory functional is the average
of the quantity $e^{i\phi}$ with this distribution. But can the full
distribution function be compared to experiment? In the first set of
experiments \cite{Nakamura} only averages could be measured, but
recently there has been demonstrated single shot readout of the qubit
state.\cite{astafiev} Is it possible to extract the full distribution
function from such data? Unfortunately, the answer is no for the
following reason.  For each realization of the experiment, there is a
certain realization of the random external noise. This gives the phase
difference $\phi$ a certain well defined value. The next realization
gives another realization of the noise process and another value for
the phase difference and so on. Since we have no knowledge about which
realization of the noise is relevant to a certain experiment, we have
to describe the final state of the qubit by a density matrix
representing a mixed state. Since all outcomes of all possible
experiments can be calculated from the density matrix, no more information 
about the system can be learned that what is contained in the density 
matrix. Since the phase distribution was calculated ignoring the 
relaxation ($T_1$) processes the density matrix will be represented by a 
vector in the equatorial plane of the Bloch sphere (but for a mixed state it 
will not be on the surface of the sphere but at some interior point), and 
the components of this vector are exactly the real and imaginary parts of 
the phase memory functional. Thus, the phase memory functional contains all 
information that we can extract about the qubit through experiments. 
A different way to express the same is that there are many different phase
distribution  functions yielding the \textit{same} qubit density matrix, and 
there is no way that one can experimentally distinguish these in an
experiment with a single qubit. 

\subsection{Non-Gaussian $T_1$?}

Notice that it seems from the figures~\ref{f3} and \ref{f4}  that 
the energy relaxation time $T_1$ is always well described by the 
Gaussian approximation. Can we understand this in a better way? 
The Gaussian approximation is good provided the correlation time of the 
noise is much shorter than the decay time, $T_1$ or $T_2$. In that case 
separation of timescales enables one to deduce an exponential decay with 
rates:\cite{NMRbok}
\[
1/T_1 \propto S(E_0), \quad
1/T_2= 1/2T_1 +\Gamma_2^{ad}, \ \Gamma_2^{ad}\propto S(0)
\]
where $S(\omega)$ is the noise power spectrum. 
For the telegraph process with switching rate $\gamma$ and coupling
strength $v$  we find that the time correlation is given by 
\[
\langle \chi(t)\chi(0)\rangle = e^{-2\gamma |t|}
\]
so the correlation time is $1/2\gamma$ and the noise power spectrum
\[
S(\omega) \propto \frac{\gamma v^2}{\gamma^2+\omega^2}\, .
\]
This gives the relative rates 
\[
\frac{1}{T_1 \gamma} \propto \frac{v^2}{\gamma^2 + E_J^2} 
\leq \left(\frac{v}{E_J}\right)^2, \quad
\frac{\Gamma_2^{ad}}{\gamma} \propto \left(\frac{v}{\gamma}\right)^2\, .
\]
We see that as long as the noise is weak compared to the qubit splitting,
$v < E_0$, the condition $\gamma T_1 \gg 1$ for the validity of the Gaussian
approximation, is always satisfied. For the $T_2$ the situation is 
different, and the condition $\Gamma_2 < \gamma$ is violated if 
$v > \gamma$, leading to non-Gaussian behavior~\cite{galperin}. 
So in the case of dephasing the Gaussian approximation so to speak predicts
its own breakdown whereas for energy relaxation it is consistent as long 
as the noise is weak compared to the level spacing.

\acknowledgments
This work was partly supported by the U. S. Department of Energy Office of
Science through contract No. W-31-109-ENG-38.

\appendix

\section{Calculation of distribution function}\label{a1}
\subsection{Single fluctuator}
To explain the method we re-derive the result for one fluctuator
with linear coupling since this is the simplest case. Let us for
simplicity also assume that $\beta (t) =1$ at $t>0$ and start with
the calculation of $\Psi_1(t)$.

 Let us discretize the integral (\ref{phi2-t}) for $\lambda_{12}=0$
 introducing small time steps $\tau \equiv t/N$, where $N \gg 1$.
 Then the random phase shift $\phi(t)$ can be expressed as 
$$\phi(t)=\nu \tau \sum_{n=1}^N\chi_n\, , \quad
\chi_n \equiv \chi(n\tau)\, . $$ 
Hence, the integration over time can then be thought of as a random
walk process, where at each time step the random walker moves a step
$\sigma=\tau \nu/2$ in the direction depending on the
current position of the fluctuator. The steps are correlated, but
only with the previous step.  The probability that a step is in
the same direction as the previous one is
$\alpha=1-\gamma\tau$ and the probability for a step to be in
the opposite direction is $\beta=\gamma\tau$. Let $m$ be the
number of steps from the origin (so that the position is $x=\sigma
m$). We want to find the probability $P_n(m)$ to be in position
$m$ at time step $n$ (dimensional time $t_n= n\tau$). This is found by the
following method. We split the probability in two parts: the
probability to reach point $m$ coming from the right, $A_n(m)$ and
from the left $B_n(m)$, so that $P_n(m)=A_n(m)+B_n(m)$. We then
have the equations
\begin{equation}
\begin{split}
  A_{n+1}(m) &= \alpha A_n(m-1) + \beta B_n(m-1)\, ,\\
  B_{n+1}(m) &= \beta A_n(m+1) + \alpha B_n(m+1)\, .
\end{split}
\end{equation}
We need the continuum limit, letting $N \to
\infty$, and $\tau \to 0$ with $N\tau=t$ fixed.
Writing
\begin{equation}
\begin{split}
  a(\phi,t) &= a(m\sigma,n\tau) = A_n(m)\, ,\\
  b(\phi,t) &= b(m\sigma,n\tau) = B_n(m)
\end{split}
\end{equation}
and expanding to first order in $\tau$ we get
\begin{equation}
\begin{split}
  a + \tau a_t &= \alpha(a-\sigma a_\phi) +\beta(b-\sigma b_\phi) \, ,\\
  b + \tau b_t &= \beta(a+\sigma a_\phi) +\alpha(b+\sigma b_\phi)\,
  .
\end{split}
\end{equation}
Here subscripts $\phi$ and $t$ denote partial derivatives with
respect to $\phi$ and $t$, respectively. Adding and subtracting
these we get (with $p=a+b$ and $q=a-b$)
\begin{equation}
\begin{split}
  \tau p_t &= (\beta - \alpha)\sigma q_\phi \, , \\
  q + \tau q_t &= (\alpha - \beta)q - \sigma p_\phi\, .
\end{split}
\end{equation}
Differentiating the second of these and inserting $q_\phi$ from
the first we obtain the final equation for $p$
\begin{equation}\label{telegraph}
  p_{tt}+2\gamma p_t = \left(\frac{\nu}{2}\right)^2p_{\phi \phi}
\end{equation}
which is called the telegraph equation. We guess the solution
$p=e^{i(k\phi-\omega t)}$ and get the dispersion relation
\begin{equation}
  \label{eq:de1}
  \omega_\pm = -i\gamma\pm
\sqrt{\left(\frac{\nu}{2}\right)^2 \kappa^2-\gamma^2}\, .
\end{equation}
The
general solution is then
\begin{equation}\label{gensol}
  p(\phi,t) = \int_{-\infty}^\infty \frac{d\kappa}{2\pi} \left[a_\kappa
           e^{-i\omega_+t}+b_\kappa e^{-i\omega_-t} 
           \right]e^{i\kappa \phi} \, .
\end{equation}
The coefficients $\{a_\kappa ,b_\kappa \}$ can be obtained from the
initial conditions, 
\begin{eqnarray}
  \delta(\phi)&=& p(\phi,0) =
  \int_{-\infty}^\infty d\kappa \left[a_\kappa +b_\kappa
  \right]e^{i\kappa \phi}\nonumber
  \\
   && \implies a_\kappa  + b_\kappa  = 1\, , \\
\nu\chi_0\delta'(\phi) &= &p_t(\phi,t) =
   -i \int_{-\infty}^\infty
   d\kappa \left[\omega_+a_\kappa +\omega_-b_\kappa \right]e^{ikx}\nonumber \\
    &&\implies \omega_+a_\kappa  + \omega_-b_\kappa  = -\kappa \nu\chi_0\, ,
\end{eqnarray}
which yield
\begin{equation}
\{a_\kappa ,b_\kappa \}= \frac{1}{2}\mp\frac{\kappa \nu\chi_0-i\gamma}
    {2\sqrt{\kappa^2\nu^2/4-\gamma^2}}\, .
\end{equation}
This is to be inserted into Eq.~(\ref{gensol}). 
The result is given by Eq.~(\ref{eq:df001}).

The memory functional can be expressed as the expectation value
$\Psi_1 = \int d\phi\, p(\phi ,t)\, e^{i\phi}$. Thus
\begin{equation}
\begin{split}
  \Psi_1& =  \int_{-\infty}^\infty d\kappa \left[a_\kappa e^{-i\omega_+t}
                +b_\kappa 
e^{-i\omega_-t}\right]\delta(\kappa +1)\\
     &= \frac{1}{2\mu}e^{-\gamma t} \sum_{\pm}
        \left(\mu\pm1 \pm\frac{i\nu\chi_0}{\gamma}\right)e^{\pm\mu\gamma t}
\end{split}
\end{equation}
where $\mu=\sqrt{1-\nu^2/4\gamma^2}$. This agrees with the
result of Ref.~\onlinecite{galperin}.

\subsection{Two fluctuators}
Now we turn to the case of two fluctuators.
Again we discretize (\ref{phi2-t}) to get
$$\phi_2=\tau
\sum_{n=1}^N(\nu\chi_1+\nu\chi_2+2\lambda\chi_1\chi_2)\, .$$
There are now three kinds of steps, depending on the settings of
the fluctuators. If both have the value $+\frac{1}{2}$ there is a
step $\bar{\alpha}=\nu+\frac{\lambda}{2}$, if both are $-\frac{1}{2}$ there
is a step $\bar{\beta}=-\nu+\frac{\lambda}{2}$, and for one of each the
step is $\bar{\gamma}=-\frac{\lambda}{2}$. We have the following probabilities
for each jump, depending on the previous state:
\begin{equation*}
\mbox{Previous}\left\{\begin{array}{l|ccc}&\bar{\alpha}&\bar{\gamma}&\bar{\beta}\\\hline 
\bar{\alpha}& \alpha^2&2\alpha\beta&\beta^2\\
\bar{\gamma}& \alpha\beta&\alpha^2+\beta^2&\alpha\beta\\
\bar{\beta}&\beta^2&2\alpha\beta&\alpha^2
\end{array}
\right.
\end{equation*}
The total probability must now be split in 3 parts
$P_n(\phi)=A_n(\phi)+B_n(\phi)+C_n(\phi)$, where $A_n(\phi)$ is the
probability to reach point $\phi$ at time step $n$ with a
$\bar{\alpha}$ jump, $B_n(\phi)$ with a
$\bar{\beta}$ jump and $C_n(\phi)$ with a
$\bar{\gamma}$ jump.  We have then the set of
equations
\begin{widetext}
\begin{equation}
\begin{split}
A_{n+1}(\phi) &= \alpha^2
A_n(\phi-\bar\alpha\tau)+\beta^2B_n(\phi-
\bar\alpha\tau)
   +\alpha\beta C_n(\phi-\bar\alpha\tau)\, ,\\
B_{n+1}(\phi) &= \beta^2
A_n(\phi-\bar\beta\tau)+\alpha^2B_n(\phi-
\bar\beta\tau)
   +\alpha\beta C_n\left[\phi-\bar\beta\tau\right]\, ,\\
C_{n+1}(\phi) &= 2\alpha\beta
A_n\left(\phi-\bar\gamma\tau\right)+2\alpha\beta
B_n\left(\phi-\bar\gamma\tau\right)
   +(\alpha^2+\beta^2)C_n\left(\phi-\bar\gamma\tau\right)\, .
\end{split}
\end{equation}
\end{widetext}
Again we introduce continuous variables $a,b,c$ and expand to
first order in $\tau$ to get
\begin{equation}
\begin{split}
 a_t &= -2\gamma a-\bar\alpha a_\phi+\gamma c\, ,\\
 b_t &= -2\gamma b-\bar\beta b_\phi+\gamma c \, ,\\
 c_t &= 2\gamma a+2\gamma b-2\gamma c-\bar\gamma c_\phi \,
 .
\end{split}
\end{equation}
This can be written in matrix form
\begin{equation}
{\bf a}_t = M{\bf a}, \qquad {\bf a}=
\left(\begin{array}{c}a\\b\\c\end{array}\right)  
\end{equation}
where
\begin{equation}
 M=\left(\begin{array}{ccc}
    -2\gamma -\bar\alpha\partial_\phi&0&\gamma  \\
    0&-2\gamma -\bar\beta\partial_\phi&\gamma \\
    2\gamma &2\gamma&-2\gamma +\bar\gamma\partial_\phi
\end{array}\right)  \,.
\end{equation}
We then guess the solution in the form
\begin{equation}
{\bf a}={\bf A} e^{i(\kappa \phi-\omega t)}
\end{equation}
which gives the eigenvalue equation 
\begin{equation}
-i\omega{\bf a}
  =\tilde{M}{\bf A}
\end{equation}
where $\tilde{M}$ is the matrix $M$ with $\partial_\phi$ replaced by $i\kappa$.
{}From this we get the dispersion equation for $\omega$ in terms
of $\kappa$:
\begin{eqnarray}
&&\omega^3+\left(6\gamma i-\kappa \frac{\lambda}{2}\right)\omega^2
\nonumber \\ &&
 +\left(-\kappa^2\nu^2-8\gamma^2-4\gamma
 i\kappa \frac{\lambda}{2}
 -\kappa^2\frac{\lambda^2}{4}\right)
    \omega
    \nonumber \\ &&
  -2\gamma i
  \kappa^2\left(\nu^2+\frac{\lambda^2}{4}\right)-\kappa^3\frac{\lambda}{2}
\left(\nu^2    
     -\frac{\lambda^2}{4}\right)
      =0 \, . \nonumber
\end{eqnarray}
This equation has the three solutions: $\omega_0$ and $\omega_\pm$
where $\omega_0$ is the solution that goes continuously to
$-2i\gamma$ when $\lambda\to0$. The general solution is then
\begin{equation}
{\bf a} = \int \frac{d\kappa}{2\pi} \,  \tilde{A}_\kappa 
\left(\begin{array}{c}e^{-i\omega_0t}\\e^{-i\omega_+t}\\e^{-i\omega_-t}
  \end{array}\right)e^{i\kappa \phi} \, ,
\end{equation}
where 
\begin{equation}
\tilde{A}_\kappa=\left(\begin{array}{ccc}a_\kappa^a&b_\kappa^a&c_\kappa^a\\
      a_\kappa^b&b_\kappa^b&c_\kappa^b\\a_\kappa^c&b_\kappa^c&c_\kappa^c\end{array}\right)
\end{equation}
is a matrix of coefficients that has to be determined by the initial 
conditions. 

Since we have three coefficients to determine for each of the $a$,$b$ and $c$
it appears that we need to specify both the functions $a(\phi,0)$... and the 
first two derivatives. However, because of the special form of the equations
we can calculate all derivatives at $t=0$ from the functions $a(\phi,0)$...
\begin{equation}
{\bf a}_t(\phi,0) 
    = \tilde{M}{\bf a}(\phi,0),
\qquad  {\bf a}_{tt}(\phi,0) 
    = \tilde{M}^2{\bf a}(\phi,0)\, .
\end{equation}
The typical initial conditions would then correspond to specifying the
initial type of jump in the random walk. For example if this was of
type $A$ we would have $a(x,0)=\delta(x)$ and $b(x,0)=c(x,0)=0$. Note that 
in this simple case where only one of the $a$,$b$ and $c$ are nonzero at 
$t=0$ the procedure could be simplified by writing the general solution 
for $p=a+b+c$ and initial conditions for this. The more general case would be 
that all of $a$,$b$ and $c$ are nonzero and the complete matrix $A_\kappa$ is 
needed. This would be the case for example when calculating echo signals, where
the equations after the echo pulse has to be solved with initial conditions
of this type corresponding to the solution of the equations before the echo 
pulse is applied.

Introducing the Fourier transformed functions (in $\phi$)
\begin{equation}
  \tilde{\bf a}_\kappa = \left(\begin{array}{c}\tilde{a}_\kappa\\\tilde{b}_\kappa\\\tilde{c}_\kappa\end{array}\right)
    =\tilde{A}_\kappa\left(\begin{array}{c}e^{-i\omega_0t}\\e^{-i\omega_+t}\\e^{-i\omega_-t} 
  \end{array}\right)
\end{equation}
we can write the initial conditions as
\begin{widetext}
\begin{equation}
 \tilde{\bf a}_\kappa(t=0)  =
 \tilde{A}_\kappa\left(\begin{array}{c}1\\1\\1\end{array}\right), 
 \qquad\tilde{M}
\tilde{\bf a}_\kappa(t=0)
  = -i A_\kappa\left(\begin{array}{c}\omega_0\\\omega_+\\\omega_-
  \end{array}\right),
  \qquad\tilde{M}^2
\tilde{\bf a}_\kappa(t=0)
  = - \tilde{A}_\kappa\left(\begin{array}{c}\omega_0^2\\\omega_+^2\\
 \omega_-^2
  \end{array}\right)\, .
\end{equation}
These can be written more compactly if we introduce the matrix $\tilde{M}_a$ 
with the left hand sides of the above equations as columns
\begin{equation}\label{matrise}
\underbrace{
\left(
\begin {array}{c|c|c} 
\tilde{\bf a}_\kappa(t=0)
&\tilde{M}
\tilde{\bf a}_\kappa(t=0)
&\tilde{M}^2
\tilde{\bf a}_\kappa(t=0)
\end {array}
\right)}_{\tilde{M}_a} 
= A_\kappa\underbrace{\left(
\begin{array}{ccc}1&-i\omega_0&(-i\omega_0)^2\\
1&-i\omega_+&(-i\omega_+)^2\\1&-i\omega_-&(-i\omega_-)^2
\end{array}
\right)}_\Omega 
\end{equation}
\end{widetext}
From which the coefficients are found as $A_\kappa=\tilde{M}_a\Omega^{-1}$.

The final solution is then  
\begin{equation}
  p(\phi,t) = \int_{-\infty}^\infty \! \! \frac{d\kappa}{2\pi}\, \left[a_\kappa
           e^{-i\omega_0t}+b_\kappa e^{-i\omega_+t}
           +c_\kappa e^{-i\omega_-t}\right]e^{i\kappa\phi}  \nonumber
\end{equation}
where $a_\kappa=\sum_ia_\kappa^i$ and similarly for $b_\kappa$ and $c_\kappa$.
Again the average is calculated from
\begin{eqnarray*}
  \langle e^{i\phi}\rangle &=& \int d\phi e^{i\phi}p(\phi,t)
  \\ &=&a_{-1} e^{-i\omega_0t}+b_{-1}e^{-i\omega_+t}
           +c_{-1}e^{-i\omega_-t}
\end{eqnarray*}
where $\kappa=-1$ in the $\omega$ in the last expression because of the
$\delta$-function from the $\phi$ integral.

Let us find the explicit expressions for the coefficients $a_\kappa$,
$b_\kappa$ 
and $c_\kappa$ in this case. Adding the lines in Eq.~(\ref{matrise}) we get 
\[ 
\begin{split}
   a_\kappa+b_\kappa+c_\kappa &= 1\, ,  \\
 a_\kappa\omega_0+b_\kappa\omega_++c_\kappa\omega_- &=
\kappa(\nu(\chi_1^0+\chi_2^0)+
      2\lambda\chi_1^0\chi_2^0) \equiv A_\kappa \\
 a_\kappa\omega_0^2 +b_\kappa\omega_+^2+c_\kappa\omega_-^2 &=
\kappa^2(\nu(\chi_1^0+\chi_2^0)
  +2\lambda\chi_1^0\chi_2^0)^2
 \\
  -    2 & i\gamma  \kappa\nu  (\chi_1^0 +\chi_2^0)
      - 8i\gamma \kappa\lambda\chi_1^0\chi_2^0 \equiv B_\kappa\, .
\end{split}
\] 

Here $\chi_{1,2}^0$ represent the initial state of the fluctuators. Also of 
interest are the values of these averaged over the initial states of the 
fluctuators. Assuming all 4 settings are equally probable we have
\[
 A_\kappa^{av} = 0 \, , \quad
 B_\kappa^{av} = \frac{1}{2}\kappa^2(\nu^2+\frac{1}{2}\lambda^2)\, .
\]
In terms of these the coefficients are expressed as 
\begin{equation}
\begin{split}
  c_\kappa &= \frac{(\omega_0-A)(\omega_0+\omega_+)-(\omega_0^2-B)}
     {(\omega_0-\omega_-)(\omega_+-\omega_-)}\, ,\\
  b_\kappa &= \frac{-(\omega_0-A)(\omega_0+\omega_-)+(\omega_0^2-B)}
     {(\omega_0-\omega_+)(\omega_+-\omega_-)}\, ,\\
  a_\kappa &= 1-b_\kappa-c_\kappa \, .
\end{split}
\end{equation}

\subsection{General case}
The above method is in principle simple to generalize to any number
of fluctuators, but the number of equations increases exponentially in 
the number of fluctuators.

The general equation is 
\[
{\bf a}_t=M{\bf a}, \qquad {\bf a}
     =\left(\begin{array}{c}a\\b\\c\\\vdots\end{array}\right)
\]
Guessing the solution ${\bf a}={\bf A}e^{i(\kappa\phi-\omega t)}$ we get
the dispersion equation $-i\omega{\bf a}=\tilde{M}{\bf A}$, which determines
the $n$ eigenvalues $\omega_i(\kappa)$ ($i=1\ldots n$) as functions of
$\kappa$.  
Here $n=2^N$ with $N$ the number of fluctuators. The general solution is 
\[
 {\bf a} = \int \frac{d\kappa}{2\pi}  e^{i\kappa \phi} A_\kappa
  e^{{\bf \omega}t},  
  \qquad  e^{{\bf \omega}t}= 
 \left(\begin{array}{c}e^{-i\omega_1t}\\e^{-i\omega_2t}
   \\\vdots\end{array}\right)
\]
Defining
\[
 \tilde{\bf a}_\kappa 
  =
  \left(\begin{array}{c}\tilde{a}_\kappa\\\tilde{b}_\kappa\\\vdots\end{array}\right)  
  = A_\kappa e^{\bf \omega}, \mbox{   and   } 
  \tilde{M}=M(\partial_\phi\rightarrow i\kappa)
\]
we determine the coefficient matrix $A_\kappa$ from 
\begin{widetext}
\[
\underbrace{
\left(
\begin {array}{c|c|c|c} 
\tilde{\bf a}_\kappa(t=0)
&\tilde{M}
\tilde{\bf a}_\kappa(t=0)
&\cdots
&\tilde{M}^{n-1}
\tilde{\bf a}_\kappa(t=0)
\end {array}
\right)}_{\tilde{M}_a} 
= A_\kappa\underbrace{\left(
\begin{array}{cccc}1&-i\omega_1&\cdots&(-i\omega_1)^{n-1}\\
\vdots&\vdots&&\vdots\\1&-i\omega_n&\cdots&(-i\omega_n)^{n-1}
\end{array}
\right)}_\Omega 
\]
\end{widetext}
The matrix $\tilde{M}_\kappa$ can be written as the contraction of a
third order  
tensor $\tilde{M}_T$ with $\tilde{a}_\kappa$. Writing the tensor
indices we have  
\[
 [\tilde{M}_a]_{ij} = [\tilde{M}_T]_{ijk}[\tilde{\bf a}_\kappa(0)]_k, \qquad
 [\tilde{M}_T]_{ijk} = [\tilde{M}^{j-1}]_{ij}
\]
We then get 
\[
 [A_\kappa]_{ij} = [\tilde{M}_T]_{imk}[\Omega^{-1}]_{mj}[\tilde{\bf
  a}_\kappa(0)]_k 
  \equiv [M_\Omega]_{ijk}[\tilde{\bf a}_\kappa(0)]_k
\]
and 
\[
\begin{split}
 [\tilde{\bf a}_\kappa(t)]_i &= [A_\kappa]_{ij}[e^{{\bf \omega}t}]_j 
  = [T_\kappa]_{ik}[\tilde{\bf a}_\kappa(0)]_k, \\
  [T_\kappa]_{ik} &= [M_\Omega]_{ijk} [e^{{\bf \omega}t}]_j \, .
\end{split}
\]
One can 
also write equations for echo experiments. Consider the situation where 
we initially prepare a state, then apply the echo pulse at time $t_e$ and then
measure the state at the time $2t_e$ when the echo signal appears (two pulse
echo). The state just before the application of the echo pulse has to be
calculated as above and then this is used as the initial state for the 
evolution after the echo (it is assumed that the duration of the echo pulse is 
short). After the echo pulse the matrix $\tilde{M}$
is changed because all jumps 
of the random walk changes sign. Then also the $\omega_i$ change. Let 
$\tilde{M}^-$, $\omega_i^-$ represent quantities before the echo pulse and 
 $\tilde{M}^+$, $\omega_i^+$ after the pulse. 
Then 
\[
\begin{split}
 [\tilde{\bf a}_\kappa(t_e)]_i &= [T_\kappa^-]_{ik}[\tilde{\bf a}_\kappa(0)]_k\\
 [\tilde{\bf a}_\kappa(2t_e)]_i &= [T_\kappa^+]_{ik}[\tilde{\bf a}_\kappa(t_e)]_k
\end{split}
\]
with 
\[
 [T_\kappa^-]_{ik} = [M_\Omega^-]_{ijk} [e^{{\bf \omega^-}t_e}]_j, \qquad
 [T_\kappa^+]_{ik} = [M_\Omega^+]_{ijk} [e^{{\bf \omega^+}t_e}]_j
\]\, .


\begin{thebibliography}{10}

\bibitem{Nakamura}
Y. Nakamura, Yu. A. Pashkin, and J. S. Tsai, Nature {\bf 398}, 786 (1999).
Y. Nakamura, Yu. A. Pashkin, T. Yamamoto, and J. S. Tsai, \prl {\bf
  88}, 047901 (2002). 
\bibitem{Devoret} 
D. Vion, A. Aassime, A. Cottet et al., Science {\bf 296}, 886 (2002).
\bibitem{paladino}
E. Paladino, L. Faoro, G. Falci, and R. Fazio, \prl {\bf 88}, 228304 (2002).
\bibitem{galperin}
Y. M. Galperin, B. L. Altshuler and D. V. Shantsev, cond-mat/0312490;
cond-mat/0511238. 
\bibitem{makhlin-review}
A. Shnirman, Y. Makhlin, and G. Sh\"on, Physica Scripta {\bf T102}, 147 (2002).
\bibitem{vanHarlingen}D. J. Van Harlingen, T. L. Robertson,
  B. L. T. Plourde, P. A. Reichardt, T. A. Crane, and John Clarke,
  cond-mat/0404307. 
\bibitem{makhlin}
Y. Makhlin, and A. Shnirman, PRL {\bf 92}, 178301 (2004).
\bibitem{lerner}
 Alex Grishin, Igor V. Yurkevich, Igor V. Lerner, cond-mat/0412377 (2004). 
\bibitem{makhlin2}
J. Schriefl, M. Clusel, D. Carpentier, P. Degiovanni, and Y. Makhlin,
cond-mat/0404641. 
\bibitem{NMRbok} B. P. Cowan, \textit{Nuclear magnetic resonance and
 relaxation}  (New York : Cambridge University Press, 1997).
\bibitem{astafiev2}
O. Astafiev, Yu. A. Pashkin, Y. Nakamura, T. Yamamoto, and J. S. Tsai,
unpublished. 
\bibitem{astafiev}
O. Astafiev, Yu. A. Pashkin, T. Yamamoto, Y. Nakamura, and J. S. Tsai,
\prb {\bf 69}, 180507 (2004). 


\end{thebibliography}
\end{document}